\documentclass[showpacs,prl,twocolumn]{revtex4}
\usepackage{graphicx}
\usepackage{epsfig}
\usepackage{amstext}
\usepackage{amssymb}

\usepackage{bm}
\usepackage{subfigure}

\begin{document}
\title{Chiral confinement in quasirelativistic Bose-Einstein condensates}

\author{M. Merkl$^{1}$, A. Jacob$^{2}$, F. E. Zimmer$^{1}$, P. \"Ohberg$^{1}$ and L. Santos$^{2}$}
\affiliation{(1) SUPA, Department of Physics, Heriot-Watt University, Edinburgh, EH14 4AS, United Kingdom}
\affiliation{(2) Institute for Theoretical Physics, Appelstr. 2, Leibniz University, Hannover, Germany}
\date{\today}

\begin{abstract}
In the presence of a laser-induced spin-orbit coupling an interacting ultra cold spinor Bose-Einstein condensate  
may acquire a quasi-relativistic character described by a non-linear Dirac-like equation. We show that as a result of the 
spin-orbit coupling and the non-linearity the condensate may become self-trapped, resembling 
the so-called chiral confinement, previously studied in the context of the massive Thirring model. 
We first consider 1D geometries where the self-confined condensates present an intriguing 
sinusoidal dependence on the inter-particle interactions. We further show that multi-dimensional 
chiral-confinement is also possible under appropriate feasible laser arrangements, 
and discuss the properties of 2D and 3D condensates, which differ 
significantly from the 1D case.
\end{abstract}
\maketitle

Although cold gases are typically neutral, artificial electromagnetism may be 
induced by several means, including rotation~\cite{Dalfovo1999}, manipulation of atoms in optical lattices~\cite{Ruostekoski2002,Jaksch2003,Osterloh2005}, and 
the use of laser arrangements~\cite{Ruseckas2005}. Interestingly, seminal experiments on optically created gauge fields 
have been recently reported~\cite{Lin2009}.
Artificial electromagnetism has attracted a growing attention 
in recent years, partially due to the possibility of achieving non-Abelian gauge fields~\cite{Osterloh2005,Ruseckas2005}, which  
establish fascinating links between cold gases and high-energy physics~\cite{Osterloh2005,Vaishnav2008,Pietila2009}. A striking example 
is given by the possibility of inducing quasi-relativistic physics in cold atoms 
despite the extremely low velocities involved~\cite{Juzeliunas2008}. In particular, under proper conditions cold atoms may experience an effective spin-orbit coupling, which leads to a Dirac cone in the dispersion~\cite{Jacob2007}, resembling the case of yet another 
paradigm of modern physics, namely graphene~\cite{Novoselov2005,Geim2007}. Similar phenomena are expected in cold atoms and graphene 
including Veselago lensing~\cite{Cheianov2007,Juzeliunas2008,Haddad2009}.

Interparticle interactions lead to inherent nonlinearities in Bose-Einstein condensates (BECs). 
At sufficiently low temperatures the BEC physics is described by a nonlinear Sch\"odinger equation similar to that 
found in nonlinear optics~\cite{Dalfovo1999}. Resemblances between both fields have been 
successfully explored in recent years, most remarkably in what concerns 
the physics of solitons~\cite{Burger1999,Denschlag2000,Khaykovich2002,Strecker2002,Eiermann2004}, for which 
nonlinearity and dispersion compensate leading to a non-dispersing solution. Nonlinearity plays also an important role 
in high-energy physics. Indeed, non-linear Dirac equations (NLDEs),  and more generally non-linear spinor fields, 
have been studied extensively, starting with the pioneering works of Ivanenko~\cite{Ivanenko1938}, Weyl~\cite{Weyl1950}, and 
Heisenberg~\cite{Heisenberg1953}. These equations 
may present also localized solutions~\cite{Soler1970,Ranada1984}.

\begin{figure}
\vspace{0.5cm}
\begin{center}
\includegraphics[width=6.5cm]{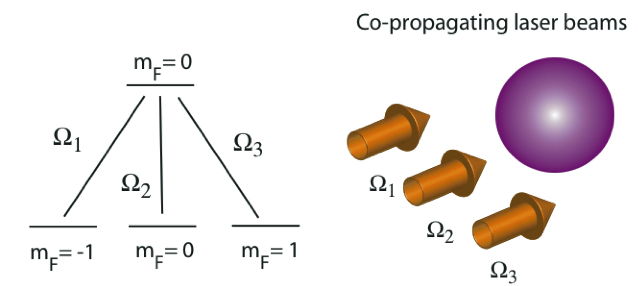}
\end{center}
\caption{One possible laser arrangement for creating a spin-orbit coupled gas, based on an atomic tripod configuration.}
\vspace*{-0.5 cm}
\label{fig:1}
\end{figure}

In this Letter we explore the nonlinear physics of a multicomponent BEC (also called spinor BEC~\cite{Ho1998}) in the presence of an optically-induced spin-orbit coupling. In the low-momentum limit, the spinor BEC may be described by a particular type of two-component NLDE. 
We show that for the case of attractive interactions, the condensate may become self-trapped, 
resembling chiral confinement, previously discussed in the context of the massive Thirring and Gross-Neveu models~\cite{Chang1975,Lee1975}. Contrary to two-component vector solitons, discussed in nonlinear optics and cold gases~\cite{Kivshar2003,Busch2001}, the two components are not only coupled by the interatomic interactions, 
but by the spin-orbit coupling as well. This leads to remarkable new features of these self-localized BECs, 
which present in 1D a peculiar sinusoidal dependence with the interaction strength. Furthermore, 
whereas with usual short-range interacting BECs solitons are stable only in 1D, we show that 2D and even 3D chiral confinement 
becomes possible by means of feasible laser arrangements, although the properties of the self-trapped BECs largely 
depend on the dimensionality of the system.


In the following we consider a Bose gas of atoms with an accessible internal tripod scheme (Fig. 1) 
formed by three ground $F=1$ states ($m_F=0,\pm 1$) and an excited $F=m_F=0$ state 
(this may be the case e.g. of the transition $5S_{1/2} (F=1) \leftrightarrow 5P_{3/2} (F=0)$ in $^{87}$Rb or the transition
$2^3S_1 \leftrightarrow 2^3P_0$ in ${^4}$He$^*$). The ground-state levels with $m_F=-1$, $0$ and $1$ are linked to the excited state 
by means of three lasers with, respectively,  $\sigma_+$, $\pi$, 
and $\sigma_-$ polarizations, Rabi frequencies $\Omega_{1,2,3}(\mathbf{r})$, and phases $S_{1,2,3}(\mathbf{r})$.
The atom-light interaction results in the appearance of two dark states $|D_{1,2}(\vec{r})\rangle$, which are linear superpositions of the three lowest bare states.
If ${\Omega}=(\sum_i \Omega_i^2)^{1/2}$ is large compared to any other energy scale, including two-photon detuning, Doppler and Zeeman shifts and interaction energy,  
we can neglect transitions out of the dark state manifold. Assuming that the atoms are loaded into this manifold we may 
express any general state as $|{\Psi(\mathbf{r},t)}\rangle = \sum_{i=1}^2 \Psi_i(\mathbf{r},t) |{D_i(\mathbf{r})}\rangle$, where 
$\Psi_i (\mathbf{r},t)$ is the wave function for atoms in $|D_i\rangle$.  
By using the Schr\"odinger equation for the tripod scheme, including the atomic motion, and projecting onto the dark state manifold,
one finds the effective Schr\"odinger equation~\cite{Ruseckas2005}
\begin{equation} \label{eq schroedinger equation effectiv1}
 i \hbar \frac{\partial}{\partial t} \vec{\Psi} = \Big[ \frac{1}{2m} \left(\mathbf{p}-\mathbf{A}\right)^2 + \mathbf{{V}}+ 
\mathbf{{\Phi}} \Big] \vec{\Psi},
\end{equation}
where $\vec{\Psi}^T =(\Psi_1,\Psi_2)$. Here $\mathbf{p}$ denotes the momentum operator and $m$ is the atomic mass. Note that the 
spatial dependence of the laser arrangement leads to an effective vector potential matrix 
(also called the Mead-Berry connection \cite{Berry1984,Wilczek1984,Mead1992}) 
${\bf A}_{nm}=i\hbar\langle D_n({\bf r})|\nabla D_m({\bf r}) \rangle$. In addition, ${\bf V}_{nm}=V_n({\bf r})\delta_{n,m}$ and ${\bf \Phi}_{nm}=(\hbar^2/2m)\langle D_n({\bf r})|\nabla B({\bf r}) \rangle\langle B({\bf r})|\nabla D_m({\bf r}) \rangle$ 
are effective scalar potential matrices, where $|B(\vec{r})\rangle$ is the so-called bright state, i.e. the linear combination of the ground levels which 
does couple with the excited state. Note that ${V_{n}}$ 
can include any detuning of the n-th laser from the resonant transition. 

The exact form of ${\bf A}_{nm}$, ${\bf \Phi}_{nm}$ and $V_n$ depends on the incident laser beams and their relative phase and intensity ratios. To obtain non-trivial gauge potentials the light fields need to be shaped carefully. There are many techniques to shape light fields in 2D, such as using  shaped apertures or algorithms to calculate the phase-hologram patterns which shape the beam in a chosen transversal plane \cite{hologram}. Shaping arbitrary light beams in 3D is more challenging but certainly possible \cite{3dshaping}. For 3D shaping holographic techniques are often used \cite{whyte2005}. 
This results in a remarkable flexibility to achieve different gauge fields (for more details see Ref.~\cite{Ruseckas2005}). In this paper 
we are particularly interested in externally inducing an effective spin-orbit coupling. This may be achieved by employing three 
co-propagating lasers along e.g. the z axis ($S_1=S_2=S_3=kz$), with constant $|\Omega_3|$ and spatially dependent transversal profiles $\Omega_1=|\Omega_3|\cos\phi(x,y)e^{ikz}$, $\Omega_2=|\Omega_3|\sin\phi(x,y)e^{ikz}$. Assuming a density modulation along $x$ such that $\phi(\mathbf{r})=\sqrt{2}\kappa x$, 
we obtain the effective spin-orbit coupling $\mathbf{A}=\hbar\kappa \hat\mathbf{\sigma}_y \vec{e}_x$, which we employ in our discussion of the 
1D localized solutions. If the density modulation has however a polar symmetry on the $xy$ plane, i.e. $\phi(\mathbf{r})=\sqrt{2}\kappa \rho$ (with 
$\rho^2=x^2+y^2$), then $\mathbf{A}=\hbar\kappa \hat\mathbf{\sigma}_y \vec{e}_\rho$, which we consider for the case of 2D chiral confinement.
Finally, setting the laser detunings $\Delta$ such that $V_1=V_2=\hbar\Delta-\hbar^2\kappa^2/2m$, and $V_3=-V_1-2\hbar\Delta$, one obtains for both 1D and 2D arrangements 
$\mathbf{\Phi}+\mathbf{V}=\hbar\Delta\hat\mathbf{\sigma}_z$. 


The spin-orbit coupling term $\mathbf{p}\cdot\bm{\sigma}$, where
$\bm{\sigma}=\hat{\sigma}_y\vec{e}_{x,\rho}$ leads to a single-atom dispersion 
law characterized by two branches $E_\pm (p)=(p^2+\hbar^2\kappa^2)/2m \pm \left ( \hbar^2\Delta^2+\hbar^2\kappa^2p_{x,\rho}^2/m^2 \right )^{1/2}$. The $E_-$ 
branch presents two minima (or a continuous ring of minima). As a result the ground-state of the many-body system may become fragmented, as recently 
discussed in Ref.~\cite{Stanescu2007}. Fragmentation (and the consequent absence of condensation) would preclude the use of the 
Gross-Pitaevskii (GP) formalism employed below for the analysis of ground state properties. The analysis is however well justified if 
we consider an initial (well defined) scalar BEC in one of the ground state levels, and adiabatically switch on the lasers, which allows a transfer of the BEC (in absence of dissipation) into the dark-state manifold.


Inter-atomic interactions do play an important role in the properties of the Bose gas at low temperatures. In the following we consider the case 
in which the interaction energy $\ll \hbar\Omega$, and hence we can still restrict ourselves to the dark-state manifold. Note that the 
ground states $|j=-1,0,1 \rangle$ constitute a spin-1 Bose gas. Short-range interactions are dominantly $s$-wave and due to symmetry may occur only in two 
different channels with a total spin of the pair $0$ and $2$, which are characterized by the corresponding $s$-wave scattering  lengths $a_0$ and $a_2$~\cite{Ho1998}. 
These scattering lengths are in principle different, although in practice they are very similar. Below, and for simplicity of the discussion, 
we consider $a_0=a_2=a$ (for $a_0\neq a_2$ the system remains in the dark-state manifold, but the equations coupling the two dark-state components are 
by far much more complicated). In the following we introduce the coupling constant $g=4 \pi \hbar^2 a/m$. 


Within the GP formalism, the interacting bosons in the dark-state manifold are described by a spinor GP-like equation with a 
spin-orbit coupling, which in its time-independent form acquires the form
\begin{equation} \label{eq-inverse-se2}
\mu\vec{\Psi}=\left[ \frac{1}{2 m} ( {\bf p}+\hbar\kappa\bm{\sigma})^2+ 
\hbar\Delta\hat\sigma_z +g\vec{\Psi}^\dag\cdot\vec{\Psi}\right]\vec{\Psi},
\label{nlse2}
\end{equation}
where $\mu$ is the chemical potential, and we consider $g<0$. For a wavepacket with $\langle p \rangle =0$ and momentum width $\Delta p \ll 2\hbar\kappa, \sqrt{2\hbar\Delta m}$ we can safely neglect the $\mathbf{p}^2$ term in Eq. (\ref{nlse2}). This results in a nonlinear Dirac-like equation, 
\begin{equation} \label{eq-nonlinear-Dirac}
\varepsilon\vec\Psi = \left[\mathbf{p}\cdot{\bm\sigma}+\hat\sigma_{z}
-\gamma\vec{\Psi}^\dag\cdot\vec{\Psi}\right]\vec{\Psi}
\end{equation}
where $\varepsilon=\mu/\hbar\Delta$, $\gamma=|g|/\hbar\Delta l_0^{d}$ for a dimension $d$, and we have employed as length unit $l_0=\hbar\kappa/m\Delta$. As discussed below, the solutions which follow
have a typical width $\gtrsim l_0$. As a consequence, the neglection of the
$p^2$ term above is valid only for sufficiently small detunings $\Delta \ll 2\hbar \kappa^2/m$.

As already mentioned NLDEs have being extensively studied in high-energy physics. Interestingly, it has been shown, in the context of chiral confinement in the 
massive Thirring and Gross-Neveu models, that Eq.~(\ref{eq-nonlinear-Dirac}) supports in the 1D case an exact self-localized solution~\cite{Chang1975,Lee1975}. 
In the following we show that these 1D self-trapped solutions present remarkable properties, which have not been explored 
in the high-energy context due to the limitations of the original physical models. In addition, we shall show that multi-dimensional 
chiral confinement is also possible with feasible laser arrangements, although it differs significantly from the 1D case.


We study first a 1D scenario along the $x$-direction, assuming a sufficiently strong transversal 
harmonic confinement (with frequency $\omega_\perp$) in the $yz$ plane, such that $\hbar \omega_\perp \gg \mu$ (we assume that the temperature 
is also much lower than the transversal oscillator energy). We consider that the system experiences a spin-orbit coupling (see above) 
$\mathbf{p}\cdot\mathbf{\sigma}=p_x \hat{\sigma}_y$. Using $\vec\Psi=\eta (\cos\varphi,sin\varphi)^T$ Eq.~(\ref{eq-nonlinear-Dirac}) 
transforms into two coupled equations:
\begin{eqnarray}
\frac{d\varphi}{dx}&=&\varepsilon-\cos(2\varphi)+\gamma\eta^2\label{nldt1}\\
\frac{d\eta}{dx}&=&-\eta\sin(2\varphi).\label{nldt2}
\end{eqnarray}
which lead to $\eta^2 \left ( \varepsilon^2+\gamma\eta^2/2-\cos 2\varphi \right )={\rm const}$. 
Imposing localization ($\eta\rightarrow 0$ for $x\rightarrow\pm\infty$) one obtains 
$\varepsilon+\gamma\eta^2/2-\cos 2\varphi =0$. Inserting this conservation law back in~(\ref{nldt1}) and (\ref{nldt2}) 
we obtain the localized solution \cite{Chang1975}
\begin{eqnarray}
\varphi(x)&=&\tan^{-1}(\sqrt{\beta}\tanh(\lambda x)) \label{phi} \\
\eta^2(x)&=&\frac{2 (1-\varepsilon)/\gamma}{\cosh^2(\lambda x)+\beta\sinh^2(\lambda x)}\label{soliton}
\end{eqnarray}
where we have introduced the notation $\beta=(1-\varepsilon)/(1+\varepsilon)$ and $\lambda=\sqrt{1-\varepsilon^2}$. 
From these expressions it is clear that only solutions with $|\varepsilon|<1$ (i.e. $|E|<\hbar\Delta$) are localized. 

Imposing the normalization of the 1D wavefunction $\int_{-\infty}^\infty dx \eta^2=1$ leads to the 
condition $\gamma=4\tan^{-1}\sqrt{\beta}$. As a consequence both normalization and localization conditions 
are only fulfilled for $0<\gamma< 2\pi$. In this regime the energy presents a remarkable 
sinusoidal dependence $\varepsilon=\cos(\gamma/2)$. 
Note that contrary to the high-energy case, where the constraint to positive energy 
fermion states demands $\gamma\leq\pi$~\cite{Chang1975}, in ultracold gases $\gamma>\pi$ is also possible. 
This leads to a peculiar behavior of the BEC wavefunction when $\gamma$ approaches $2\pi$. 
In the regime $0<\gamma< 2\pi$ the square width of the soliton is
\begin{equation}
\langle x^2 \rangle = \frac{\pi^2-(\gamma/2)^2}{12\sin^4(\gamma/2)},
\end{equation}
which clearly diverges both at $\gamma=0$ and $2\pi$. The divergence has however a rather different meaning in both cases. 
For $\gamma=0$ there is no localized solution, as one could expect due to the absence of interactions. On the contrary for $\gamma=2\pi$, the localized solution exists but acquires a 
Lorentzian form $~1/(1+4x^2)$, with divergent $\langle x^2\rangle$ but finite half-width at half-maximum $1/2$. Direct numerical simulations of Eq. (1) confirm the existence of these localized solutions, which remain well self-localized for $\hbar\Delta < 0.1 \hbar\kappa^2 / 2m$.

Due to the form of $\varphi(x)$, the relative distribution of the population between the two components also presents a 
peculiar dependence in $\gamma$. The chirality $\chi$ (defined as the difference between the populations 
in $|D_1\rangle$ and $|D_2\rangle$) becomes $\chi=\int dx \eta^2 \cos(2\varphi)=\frac{2}{\gamma}\sin(\gamma/2)$, which monotonously 
decreases from $1$ (all atoms in $|D_1\rangle$) at $\gamma=0$ to $0$ (equal admixture of both dark states) at $\gamma=2\pi$. The density profile of both components is also quite different, since $|D_2\rangle$ presents a node at $x=0$, whereas $|D_1\rangle$ is maximal at the center.

\begin{figure}
\vspace{0.5cm}
\begin{center}
\includegraphics[width=6.0cm]{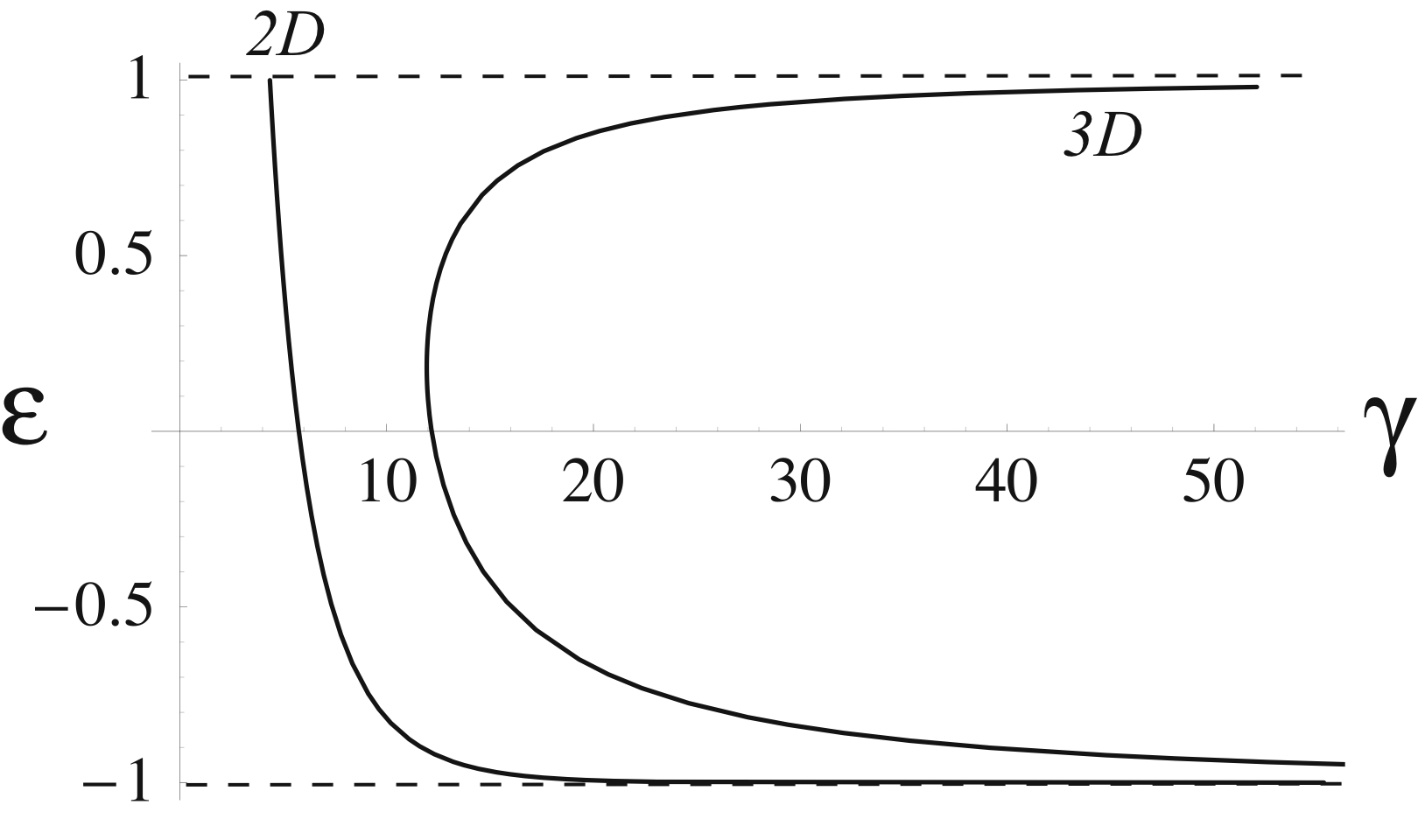}
\end{center}
\caption{Energy $\varepsilon$ as a function of the interaction strength $\gamma$ for a 2D and 3D arrangement. Note the existence of 
a solution for all $\gamma>4.36$ in 2D and a double solution for all $\gamma>11.94$ in 3D.}
\vspace*{-0.5cm}
\label{fig:3}
\end{figure}

\begin{figure}
\includegraphics[width=8.5cm]{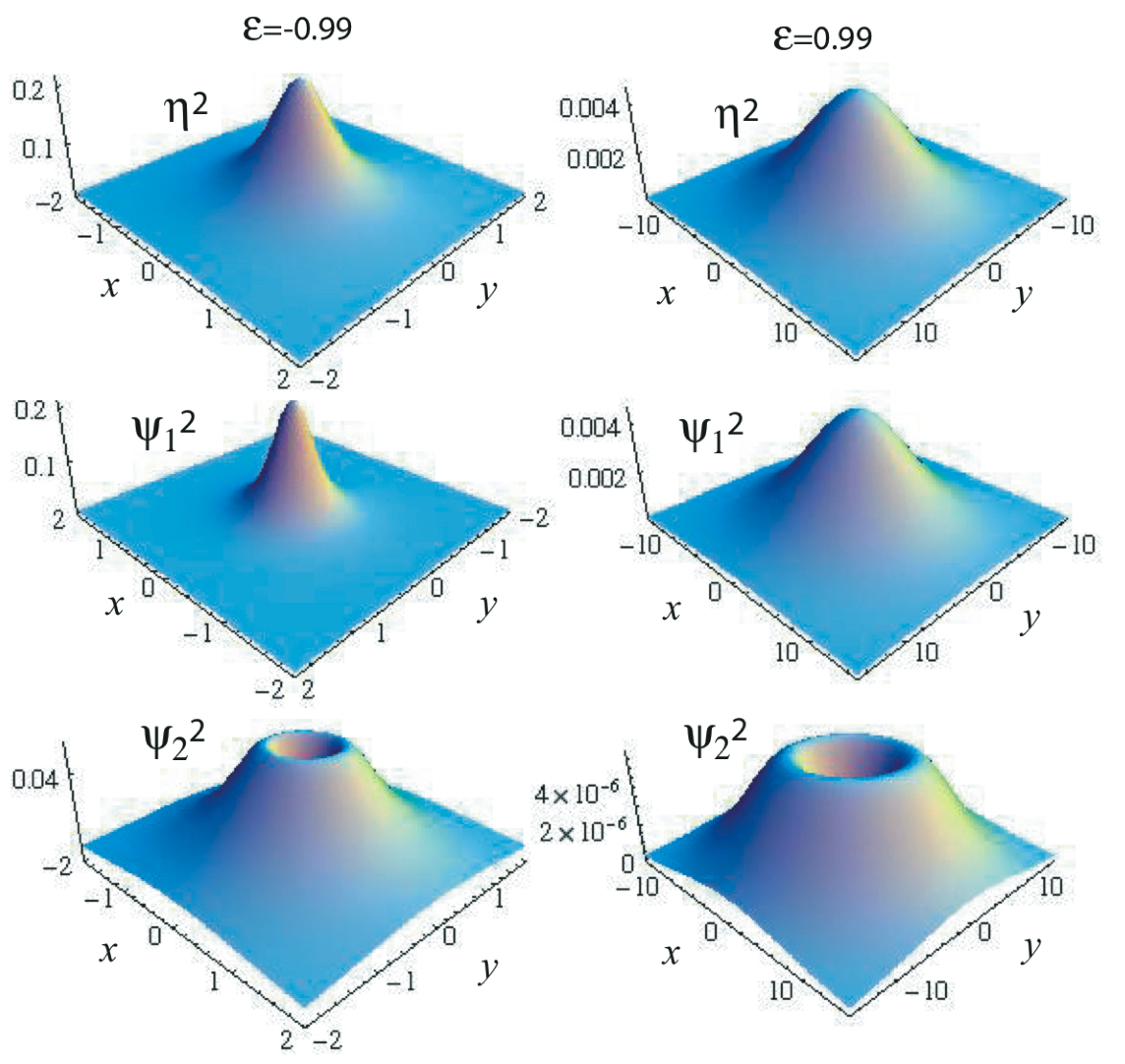}
\caption{Left column: $\varepsilon=-0.99$ and $\gamma=18.63$. The localized total density $\eta^2$ is shown at the top. The density of each component $\Psi_{1,2}^2$ are strikingly different, with component 1 forming a peak in the center. Most particles are occupying component 2. Right column: $\varepsilon=0.99$ and $\gamma=4.36$. The component 1 is forming a peak in the centre where now most particles are located. The $x$ and $y$ coordinates are in units of $l_0$ and the densities in units of $1/l_0^2$.}
\vspace*{-0.5cm}
\label{fig:2}
\end{figure}

Note that the discussed sinusoidal dependence has its origin in the 1D normalization of~(\ref{soliton}). We should hence expect a rather different 
behavior in higher dimensions. This is indeed the case. We consider in the following a 2D scenario on the $xy$ plane (we assume a strong confinement 
along $z$) in the presence of a spin-orbit coupling $\mathbf{p}\cdot\mathbf{\sigma}=p\hat\sigma_y$, where $p^2=p_x^2+p_y^2$.
Eqs.~(\ref{phi}) and~(\ref{soliton}) can be readily generalized to the 2D case by replacing $x$ by the radial coordinate $\rho$. 
The normalization of (\ref{soliton}) is however different $2\pi \int\rho \eta^2(\rho) d\rho=1$, which requires
\begin{equation}
 \gamma=4\pi \left [ \frac{s\ln 2 -L(s)}{\sin 2s}\right ], \label{gamma2D}
\end{equation}
where $\varepsilon=\cos(2s)$ ($0\leq s \leq \pi/2$) and $L(s)=-\int_0^s \ln (\cos s') ds'$ is the Lobachevsky function~\cite{GradsteynBook}. By inverting~(\ref{gamma2D}) we obtain $\varepsilon(\gamma)$~(Fig.~\ref{fig:3}). As mentioned above a localized solution requires $|\varepsilon|\leq 1$, which is only possible for $\gamma>\gamma_c=2\pi\ln 2\simeq 4.36$. 
Therefore, contrary to the 1D localization which may occur for $0<\gamma< 2\pi$, 
the 2D localized BEC requires a minimal interaction strength $\gamma_c$ at which the BEC width diverges. 
In addition both localization and normalization conditions are simultaneously fulfilled for arbitrary $\gamma>\gamma_c$. 
For increasing $\gamma>\gamma_c$, $\varepsilon$ decreases monotonously from $1$ to $-1$.  
For $\gamma\gtrsim 15$, $\varepsilon\simeq -1$ and the localized wavefunction converges to 
a Lorentzian shape $\eta^2(\rho)=8\pi\gamma^{-1}(1+4\rho^2)^{-1}$.  Fig.~\ref{fig:2} shows 2D localized BECs. Note that the behavior of the chirality $\chi$ in 2D differs from the 1D case. In particular
for $\varepsilon\rightarrow 1$, $\chi\rightarrow 1$ as in 1D, and for $\varepsilon\rightarrow -1$, $\chi\rightarrow -1$, {\it i.e.}, $|D_2\rangle$ dominates.
Note also that as in 1D the $|D_2\rangle$ state has a minimum at the center, whereas $|D_1\rangle$ is there maximal.

Finally, a similar solution may occur also in 3D if $\mathbf{p}\cdot\mathbf{\sigma}=p\hat\sigma_y$, where $p^2=p_x^2+p_y^2+p_z^2$.
In that case the 3D normalization requires that $\gamma=2s(\pi^2-4s^2)/3\sin^2(2s)$~(Fig.~\ref{fig:3}). The localized solution exists only for $\gamma>\gamma_c\simeq 11.94$. Contrary to 2D where the BEC width diverges at $\gamma_c$ and hence the wavefunction experiences a smooth crossover from localization into 
delocalization, in 3D at $\gamma=\gamma_c$, $\varepsilon<1$ and the BEC width remains finite. Hence in 3D there is an abrupt transition between localization 
and delocalization regimes. Finally, let us point out that, interestingly, for $\gamma>\gamma_c$, there are actually two localized solutions. At 
$\gamma\rightarrow\infty$, one of the solutions becomes unbound and the other a Lorentzian shaped function.


In summary, the interplay between interactions and an optically induced spin-orbit coupling may lead to self-localized 
Bose-Einstein condensates in 1D, but also in 2D and 3D. Self-localization in NLDE was previously studied 
in the context of chiral confinement in massive Thirring and Gross-Neveu models. However, for the case of cold gases 
a novel regime of parameters (and dimensionalities) are possible, allowing for 
remarkably rich physics which largely depends on the system dimensionality, ranging from a sinusoidal 
interaction dependence in 1D, to two possible self-trapped solutions in 3D scenarios. 
These results provide exciting new perspectives for the 
nonlinear physics of condensates in artificially-induced gauge fields.

\acknowledgements

This work was supported by EPSRC UK, DFG (SFB407, QUEST), and ESF (EUROQUASAR).

\end{document}